\documentclass[11pt]{article}
\usepackage{epsfig} 
\newcommand{\sect}[1]{ \section{#1} \setcounter{equation}{0} }

\newcommand{\Dslash}{D \! \! \! \! /}

\newcommand{\threehalves}{\mbox{\small{$\frac{3}{2}$}}}

\newcommand{\MSbar}{\overline{\mbox{MS}}}

\newcommand{\Nf}{N_{\!f}}

\newcommand{\NA}{N_{\!A}}

\begin{document}
\title{Low momentum propagators at two loops in gluon mass model}
\author{J.A. Gracey, \\ Theoretical Physics Division, \\ 
Department of Mathematical Sciences, \\ University of Liverpool, \\ P.O. Box 
147, \\ Liverpool, \\ L69 3BX, \\ United Kingdom.} 
\date{}
\maketitle 

\vspace{5cm} 
\noindent 
{\bf Abstract.} We compute the two loop corrections to the gluon propagator
for low momentum in a gluon mass model. This model has recently been proposed 
as an alternative to the Gribov construction in the way it handles Gribov
copies in the gauge fixing. The corrections provide improvements for estimating
the point where the gluon propagator freezes in relation to lattice data.

\vspace{-15.2cm}
\hspace{13cm}
{\bf LTH 1021}

\newpage 

\sect{Introduction.}

The low momentum properties of the propagators of gluons and Faddeev-Popov
ghosts in Quantum Chromodynamics (QCD) has been the subject of intense interest
in recent years. See, for example, \cite{1,2,3,4,5,6,7,8,9,10,11}. The main 
motivation for such studies rests in their relation to colour confinement. In a
non-abelian gauge theory gluons have a fundamental style of propagator at high 
energy indicating an effective particle interpretation. Specifically the fields
are effectively massless and asymptotically free. That such vector gauge bosons
are not observed in nature is indicative of a deeper structure from the point 
of view of the full propagator. In other words taken over the whole energy 
spectrum the gluon propagator does not have a fundamental structure with a 
simple pole at zero or a non-zero value. Instead recent lattice gauge theory 
studies have reached the consensual picture that the propagator not only does 
not have a pole but it freezes at zero momentum to a non-zero value 
\cite{1,2,3,4,5,6,7,8,9}. This positivity violating form is consonant with a
confined gluon and hence no free colourful strong force vector bosons are seen 
in nature in isolation. While the non-zero frozen value of the propagator 
clearly has a scale it is strictly speaking not a true mass since that is 
adduced from a simple pole of a propagator. However, the key point from lattice
data is that some scale can be associated with the infrared dynamics of the 
gluon. From a quantum field theory point of view understanding the origin of 
such a scale in a Lagrangian context could prove useful in gaining an insight 
into the colour confinement mechanism. 

Over the years there have been various attempts at such an understanding. For
instance, in \cite{12} a gauge independent mass term for the gluon was
considered. When quantum corrections were included this produced a gluon 
propagator whose full momentum dependence was remarkably consistent with the
lattice data of recent years. Indeed the consequences of such an underlying
gluon mass was explored in \cite{13} where the glueball spectrum was
constructed using a variational method approach. Another of the early and deep 
insights was provided by Gribov in \cite{14}. There it was observed that the 
ambiguity deriving from the non-uniqueness of fixing the gauge globally could 
induce a cutoff in the path integral defining the gluon configuration space. 
Such a restriction affected the structure of the gluon propagator but in such 
a way as to retain its massless fundamental-like behaviour at high energy. 
However, taken in the full the propagator contained a new scale parameter 
called the Gribov mass and produced a positivity violating propagator, 
\cite{14}. Though this new parameter was not an independent quantity as it 
satisfies a gap equation. In the current context the Gribov propagator suffered
from the drawback that while freezing in the infrared to a finite value, that 
finite value was actually zero and not consistent with current lattice data. A 
modification \cite{13,14,15,16} of the Gribov construction built on Zwanziger's
localization programme, \cite{19,20,21,22,23,24,25,26,27,28,29}, can be used to
model the data. Moreover, that required the condensation of the additional 
localizing fields introduced in \cite{16,17,18} to produce a dynamic mass scale
aside from the Gribov mass which was already present. While successful in 
modelling the data it does not appear to be constructed in a fundamental way 
within the ethos of Gribov's elegant construction. 

In recent years there has been a re-examination of the Gribov copy problem in a
series of articles, \cite{30,31,32}. For instance, in \cite{30} a reweighting
of the average of Gribov copies has produced a new Lagrangian in which to study
Gribov copy issues and thence to examine their effect on the propagator at low 
momentum. In essence a new type of gauge fixed Lagrangian emerges with a new 
gauge parameter which has dimensions of mass. As an aside we note that like the
work we have summarized so far we are concentrating throughout this article on 
the Landau gauge. This new gauge parameter is related to counting replicas 
derived from the Gribov copies, \cite{30}. The resulting Lagrangian bears a 
strong resemblance to the nonlinear gauge known as the Curci-Ferrari gauge, 
\cite{33,34}. Moreover, it has similarities to earlier attempts to have a copy 
free gauge fixing such as those advocated in \cite{35,36,37,38}. In those 
articles a gluon mass term was added to the Lagrangian with the mass playing a 
role similar to a gauge parameter. The fixing was copy free since the 
additional term had the property of gauge invariance. Although, like the Gribov
Lagrangian, the new Lagrangian was no longer local similar to the situation in
\cite{12}. More recently, the one loop analysis carried out for the propagators
in \cite{30} has been extended to one loop for vertex functions in three as 
well as four dimensions, \cite{32}. While such propagator and vertex functions 
derived in \cite{32} appear to be in general qualitative accord with lattice 
data, it is worth noting that parallel analyses using the Gribov-Zwanziger 
Lagrangian and its extensions also qualitatively agree with data. At this one 
loop precision it would be overambitious to hope for quantitative agreement. 
Instead one could aim to produce tests whereby certain Lagrangian motivated 
ans\"{a}tze are excluded. For instance, in \cite{39} the power corrections to 
the triple gluon vertex at one loop at the symmetric point had dimension two or
four as the leading corrections depending on whether they were determined in a 
Gribov-Zwanziger or refined Gribov-Zwanziger Lagrangian compared with a gluon 
mass motivated model akin to \cite{30,31,32}. Lattice data has not currently 
reached enough accuracy in the intermediate energy range where such power 
corrections can be differentiated between. Nevertheless it is worth pursuing 
this programme to higher loop order which is the purpose of this article. 

Several additional comments are in order for balance to contrast the differing 
positions between the Gribov-Zwanziger approach and that of gluon mass model. 
By the latter we include both the original Curci-Ferrari model as well as the 
copy free observation of \cite{30,31,32}. For instance, if one considers the 
energy-momentum tensor trace in the gluon mass case it is non-zero unlike the 
Gribov-Zwanziger situation. This is because the horizon condition defining the 
first Gribov region defines a gap equation for the Gribov mass which when 
imposed removes the non-zero contribution to the energy-momentum tensor trace. 
Such a gap equation derives naturally from the no pole condition of the Gribov 
construction which is not the case for a gluon mass model. Though in \cite{40} 
it was argued that one could define a gap equation consistent with the 
renormalization group equation so that the mass was not a free parameter 
similar to the Gribov case. While such a condition is not fully similar to that
of \cite{30,31,32} it is indicative that a condition of some sort is required 
for such mass models. Another issue which is not unnatural to raise concerns 
whether the gluon propagator or correlator as measured on the lattice is 
actually the same quantity as is computed in a Lagrangian approach. Using field
theories with explicit spin-$1$ mass terms could be regarded as merely 
modelling the lattice data. We include the refined Gribov-Zwanziger extension 
of \cite{16,17,18} in these comments where localizing ghost mass is included. 
Thus there is no fully accepted explanation of the underlying mechanism behind 
a frozen gluon propagator. One point of view is that the loss of Lorentz 
symmetry using lattice regularization means it is difficult to be certain of 
true zero momentum behaviour in the restoration of the continuum limit. In a 
dimensional regularization translational invariance is not lost. Equally the 
zero momentum values of the gluon and Faddeev-Popov ghost propagators are not 
renormalization group invariants. So the specific value is not protected by 
that principle. 

Instead an alternative position is that it could actually be the case that at 
low momentum the gluon propagator does indeed vanish analytically and the 
lattice definition of the spin-$1$ adjoint field in the infrared is not that of
the continuum. As an aside while there is no symmetry principle as such which 
implies a vanishing gluon propagator, arguments have been given in \cite{20,41} 
that this is the case in the Gribov-Zwanziger Lagrangian. Currently, for the
gluon this has been verified only at one loop and part of the motivation behind
this article is actually to provide the computational algorithm for a future  
Gribov-Zwanziger analysis at two loops. Recently some evidence has been 
provided in \cite{42} that the lattice gluon variables may not be the same as 
those of the continuum. To understand this we need to recall properties of the 
Gribov-Zwanziger formalism. In the pure case \cite{14} the gluon propagator 
vanishes in the infrared with the Faddeev-Popov ghost propagator enhancing. The
lattice finds a frozen gluon propagator and a non-enhanced Faddeev-Popov ghost 
propagator. The latter behaviour can be accommodated non-uniquely in the 
refined Gribov-Zwanziger scenario, \cite{16,17,18}. However, in both scenarios 
there are predictions for the bosonic localizing ghost propagator. In the pure 
case it enhances like the Faddeev-Popov ghost which has been examined at one 
loop in \cite{18,43}. An all orders analysis was also provided in \cite{41}. 
However, in the refined case at low momentum the bosonic localizing ghost 
propagator can either freeze to a non-zero value or behave in a non-enhanced
way, \cite{18}. The actual property depends on the different condensation 
channels. What has been interesting in the recent lattice study of the bosonic 
localizing ghost propagator of \cite{42} is that an {\em enhanced} propagator 
emerges for it. In any Lagrangian analysis no such enhancement is present when 
the gluon propagator freezes to a non-zero finite value. Instead the vanishing 
of the gluon propagator at zero momentum is inextricably linked to the 
enhancement of its ghost partners. While \cite{42} represents an early lattice 
investigation into the properties of this propagator it is perhaps premature to
make a final statement on the difference in definition of gluon variables in 
various approaches. 

Returning to our focus here we will extend the propagator analysis in the gluon
mass model of \cite{30} to two loops as the initial stage in providing more 
accurate tests. However, at this loop order this inevitably has to be in a 
limit since the underlying full two loop basic master Feynman integrals have 
yet to be determined as closed functions of mass for all values of momentum. 
Instead we will perform the analysis using a zero momentum expansion of the 
propagators within a Feynman integral. Such a computation is possible due to 
the algorithm developed in \cite{44}. One aim is to refine the value of the 
propagator at zero mass. As discussed in \cite{30} while the tree propagator of
the model is similar to that of a scalar field the one loop corrections wash 
out the pole and produce a positivity violating propagator whose behaviour and 
shape is qualitatively in keeping with data, \cite{30,32}. Thus having a 
correction to the one loop freezing value of the gluon propagator in this model
will aid the extraction of the fundamental scale that appears to be associated 
with the colour confinement picture. Although we have concentrated on the gluon
a partner in the analysis is the Faddeev-Popov ghost. In the Gribov 
construction, \cite{14}, the ghost propagator was found to behave as a dipole 
propagator in the deep infrared. Hence it satisfied the Kugo-Ojima colour 
confinement criterion of \cite{45,46}. Indeed more recently, \cite{47}, this 
condition has been discussed in gauges other than Landau with indications that 
it may be pointing the way to understanding gluon confinement in a gauge 
independent fashion. Here we will also refine the two loop structure of the 
Faddeev-Popov ghost propagator and extend the computation of \cite{32}. Such an
approximate calculation of this and the gluon propagator will also serve as a 
useful independent check on any future {\em full} two loop analysis which is 
another motivation for the article.

The article is organized as follows. We provide the background quantum field
theory to our computations in the next section including the essentials of the
computational algorithm we will use. The results of its application to the four
and three dimensional models are given in sections 3 and 4 respectively.
Concluding remarks are given in section 5 while an appendix summarizes the 
expansion of several massive two loop propagator integrals.

\sect{Background.}

We begin by recalling the key features of the gluon mass model, 
\cite{30,31,32}, in the Landau gauge which are required for the two loop 
analysis. The Lagrangian is \cite{30,31}
\begin{equation}
L ~=~ -~ \frac{1}{4} G^a_{\mu\nu} G^{a \, \mu\nu} ~+~ \frac{1}{2} m^2 A^a_\mu
A^{a\,\mu} ~+~ \bar{c}^a \partial^\mu \left( D_\mu c \right)^a ~+~ 
i \bar{\psi} \Dslash \psi ~-~ 
\frac{1}{2\alpha} \left( \partial^\mu A^a_\mu \right)^2
\label{lagglm}
\end{equation}
where $A^a_\mu$ is the gluon field of mass $m$, $c^a$ is the Faddeev-Popov 
ghost, $\psi^i$ are the massless quarks and $D_\mu$ is the covariant
derivative. While quarks essentially play a passive role in relation to the 
infrared behaviour we have included them here partly for completeness but also 
because of their value in internal checks within the automatic symbolic 
manipulations we will carry out. The index $a$ runs over the range 
$1$~$\leq$~$a$~$\leq$~$\NA$ where $\NA$ is the dimension of the adjoint 
representation of the colour group. We have included the canonical gauge 
parameter $\alpha$ since a non-zero value is required to deduce the gluon 
propagator. Thereafter it is set to zero throughout as we will only be
considering the Landau gauge. Thus in this gauge the $A^a_\mu$ propagator is 
\begin{equation}
\langle A_\mu^a(p) A_\nu^b(-p) \rangle ~=~ -~ \frac{\delta^{ab}}{[p^2+m^2]}
\left[ \eta_{\mu\nu} ~-~ \frac{p_\mu p_\nu}{p^2} \right] ~.
\end{equation}
The Feynman rules for the remaining propagators are the same as those derived 
from the Lagrangian with $m$ set to zero. As noted in \cite{30} this mass
parameter derives from a replica limit and the sampling of the Gribov copies in 
the gauge fixing. The idea being that the replica limit is taken before the
Landau gauge limit and the presence of the mass should ensure that the deep
infrared regime can be probed using perturbation theory, \cite{30}. Indeed the
analysis of \cite{32} is not inconsistent with this proposal in relation to
lattice data. As the early investigations into the use of (\ref{lagglm})
foundered on the loss of unitarity several comments are in order. For instance
the Curci-Ferrari gauge and model have been examined on the lattice, 
\cite{48,49}, as it is believed to circumvent the Neuberger problem. In those 
studies it is argued that for perturbative Green's functions the massless 
limit is smooth and that the appropriate way to take the limit is in a 
l'H\^{o}pital sense. However, for (\ref{lagglm}) one should be careful in the 
point of view. Clearly the gluon propagator from (\ref{lagglm}) is of a 
fundamental form and as such it models lattice data. This does not imply that 
(\ref{lagglm}) represents the true position. Rather it should be regarded as an
effective theory which is valid in some region. In other words it could be 
regarded as the leading form of an expansion of a more general but unitarity 
theory which is as yet not known. Indeed from our understanding of the Gribov 
construction it would not be unexpected if such a theory was non-local. So the 
arguments concerning the loss of unitarity of the Curci-Ferrari model which 
were applied to the theory as a whole may not be applicable in this effective 
situation.

Next we detail our method of computation. First, we generate the relevant one 
and two loop Feynman diagrams for the $2$-point functions we are interested in 
using the {\sc Qgraf} package, \cite{50}. Overall there are $4$ one loop and 
$23$ two loop graphs for the gluon self energy. The respective numbers for the 
Faddeev-Popov ghost self energy are $1$ and $7$. In both cases these are larger
than one would normally encounter in an analysis of the structure of $2$-point 
functions in QCD because one has to allow for massive snail graphs deriving 
from the closure of two legs on a quartic gluon vertex. Ordinarily such a graph
would be zero in massless QCD. The electronic representations of the graphs are
then converted into {\sc Form} input notation where {\sc Form} is a symbolic 
manipulation language, \cite{51}. At this stage the Lorentz, colour and spinor 
indices are appended. As we are interested in the zero momentum behaviour of 
the $2$-point functions we cannot use the vacuum bubble expansion of 
\cite{52,53}. This is because this is primarily to determine the divergent 
parts of the contributing Feynman graphs in order to deduce the corresponding 
renormalization group functions. In other words the vacuum bubble expansion is
an automatic way of implementing the method of infrared rearrangement,
\cite{52,53}. Instead we use the method of \cite{44} which is a way of 
determining the correct finite part of basic Feynman integrals when one is 
expanding at low momentum. This is a non-trivial task as emphasised in
\cite{44}. The main reason for this is that if one naively expanded an integral
in powers of momentum, for certain integrals with massless propagators one
would obtain spurious infrared divergences. Moreover, if not handled 
systematically by some technique one could be led to obtain results which would
contradict the renormalizability of the underlying quantum field theory. Given 
these general observations in \cite{44}, where a robust method of expansion of 
such Feynman integrals was provided, the relevant parts of \cite{44} will be
discussed later. However, in order to be able to apply such a technique we 
first have to write the contributing Feynman integrals for each graph in a 
compact notation. 

In choosing to follow \cite{44} we concentrate for the most part on the two 
loop situation as the process for one loop graphs is similar but considerably 
simpler. We define the general two loop self-energy integral by 
\begin{eqnarray}
&& I_{m_1 m_2 m_3 m_4 m_5}(n_1,n_2,n_3,n_4,n_5) \nonumber \\
&& =~
\int_{kl} \frac{1}{[k^2+m_1^2]^{n_1} [l^2+m_2^2]^{n_2} [(k-p)^2+m_3^2]^{n_3}
[(l-p)^2+m_4^2]^{n_4} [(k-l)^2+m_5^2]^{n_5}}
\label{genint}
\end{eqnarray}
where $\int_k$~$=$~$\int d^dk/(2\pi)^d$ and $d$ is the space-time dimension. 
Our notation is similar to \cite{44} but we note that the order of our labels 
differs from that of \cite{44}. Also for the specific problem we are interested
in $m_i$~$\in$~$\{0,m\}$. Given the structure of this general two loop 
$2$-point function, it is possible to write, using a {\sc Form} module, all 
numerator scalar products of the three different momenta in terms of the 
denominator factors. In other words there are no irreducible tensors for 
(\ref{genint}). At this stage there are several ways of extracting the low
momentum behaviour. For instance, one could apply the method of \cite{44} 
directly to each of the contributing integrals but regard this as inefficient.
Instead we have chosen to apply the Laporta algorithm, \cite{54}. This is a 
technique to create integration by parts identities between all the necessary 
integrals and then solve them order by order from a level which includes the 
highest numerator power. The method then produces a series of relations between
all the integrals required for the full calculation and a relatively small set 
of base integrals which are known as the masters, \cite{54}. The expressions 
for these can be substituted when they are evaluated by methods other than 
integration by parts. In our particular case not all the two loop self energy 
master integrals are known for all mass configurations. For instance, some are 
available for cases which are relevant to standard model analyses, \cite{55}. 
Therefore at this stage we substitute the values for each master integral from 
the values obtained from applying the method of \cite{44} to the emerging 
master integrals. In order to apply the Laporta algorithm we have chosen to use
the {\sc Reduze} implementation, \cite{56}, which uses {\sc GiNaC}, \cite{57}, 
and is written in C$++$. Once the setup parameters are specified for the basic 
topologies and mass configuration, {\sc Reduze} creates a database. From this 
we can extract the relations for all the integrals relevant to the self-energy 
analysis in terms of the master integrals. Again we use the feature of 
{\sc Reduze} that allows for the output from the database to be written as a 
{\sc Form} module. Therefore, we can perform the full computation for each of 
the $2$-point functions of interest automatically. Moreover, we have carried 
this out in such a way that the three dimensional analysis can also be 
performed. This requires constructing a separate {\sc Form} module for
substituting the three dimensional master integrals. 

The next stage is the insertion of the low momentum expansion of the emergent
master integrals. These can be classified into several sets dependent upon the
number of propagators. In some cases the exact expression for such masters is
known which is the case for two and three propagator two loop self energy
integrals. We note that the Laporta algorithm \cite{54} is designed in such a 
way that only non-trivial masters arise. For instance, those integrals which 
are zero because of a massless one loop bubble never emerge in the {\sc Reduze}
output. So one two loop two propagator case is the product of two one loop
massive vacuum bubbles. Equally the three propagator case can be the basic
sunset topology with different mass configurations, or the product of a one
loop massless or massive self energy graph with a one loop massive vacuum
bubble. In these cases the low momentum values are trivial to substitute.
Though we note that in certain cases the Laporta algorithm may produce a sunset
topology with an irreducible numerator. For this and the four and five 
propagator cases we resort to the algorithm of \cite{44} which we briefly 
recall. 

In general, integrals such as (\ref{genint}) can be expanded in powers of
$p^2/m^2$ where $p$ is the external momentum. However, as noted in \cite{44} 
the naive expansion of the constituent propagators within the integral itself 
does not necessarily lead to the correct expansion. This is straightforward to 
see in a simple example. As is well known $I_{m0000}(1,1,1,1,1)$ is a finite 
integral and its exact value is known \cite{55}. However, if one naively Taylor
expands a massless propagator within $I_{m0000}(1,1,1,1,1)$ which depends on 
the external momentum $p$, then the terms of the Taylor series in $p$ beyond 
the leading one, when finally evaluated witin the context of the overall 
Feynman integral itself will have poles in $\epsilon$. These divergences are 
infrared in nature since the overall original Feynman integral was infrared 
finite. They arise due to the simple fact that expanding a $p$ dependent 
massless propagator, which is not protected by a non-zero mass, will naturally 
produce propagators which have infrared singularities. Hence this expansion 
approach cannot be used. However, in \cite{44} a method was developed where 
these infrared singularities could be systematically cancelled. It relies on 
understanding the particle thresholds within the basic graph when various mass 
configurations are present. Clearly if a $p$ dependent propagator has a 
non-zero mass then Taylor expanding it in the usual way does not introduce any 
infrared singularities. The problem arises in (\ref{genint}) when it is not 
possible to route the external momentum through the propagators from one 
external vertex to the other via a full set of {\em massive} propagators. An 
example of when this is not possible is when one of the one loop subgraphs of 
(\ref{genint}) is completely massless. In this and other instances, \cite{44} 
provided an algorithm where the naive expansion of the propagators in
(\ref{genint}) have the expansion of related but different integrals appended.
Symbolically this is given by 
\begin{equation}
I_\Gamma ~ \sim ~ \sum_\lambda \, I_{\Gamma/\lambda} \circ
{\cal T}_{\{q_i\}} I_\lambda 
\end{equation}
where $I_\Gamma$ corresponds to the integral of (\ref{genint}), $\Gamma$
corresponds to the original graph and $\gamma$ is a subgraph of $\Gamma$. The
sum is over a particular set of subgraphs. This is the set of subgraphs which
contain all the massive propagators and are one particle irreducible with 
respect to the massless lines, \cite{44}. The graph corresponding to
$\Gamma/\gamma$ is that where the subgraph $\gamma$ has been shrunk to a point.
Finally, within the sum the integral corresponding to the subgraph is first
expanded using the ${\cal T}_{\{q_i\}}$ operator, where $q_i$ are a particular
set of momenta of the subgraph $\gamma$, and which operates on the propagators.
The specific set $\{q_i\}$ depends in the mass configuration of the original 
Feynman integral. For instance, \cite{44},
\begin{equation} 
T_l \frac{1}{[(l-p)^2+m_4^2]} ~=~ \sum_{i=0}^\infty 
\frac{[2lp-l^2]^i}{[l^2+m_4^2]^{i+1}} 
\end{equation} 
where $m_4$ does not have to be non-zero here. Once the expansion of all the 
subgraphs has been carried out to the necessary order for the overall $2$-point
functions of the problem at hand, the integration over the two loop momenta of 
the integral of the master is performed. The individual integrals which emerge 
to complete the determination of each master are either two loop massive vacuum
bubbles, the product of one loop integrals or massless two loop integrals. At 
this stage we have chosen to build a second {\sc Reduze} database in order to 
handle the reduction of the first of these three classes down to a base set of 
two loop master vacuum bubbles. The explicit expressions for these have been 
known for a long time for four dimensions. For example, see \cite{58,59,60}, 
for four dimensions and \cite{61} for three dimensions. Once the expansions 
have been obtained they are substituted at the appropriate point of the overall
automatic {\sc Form} computation. We note that in developing this {\sc Form} 
module we have checked the expansion algorithm we have written for the known 
cases given in \cite{44} in four dimensions and find agreement. As a final 
check on the overall expansion, when all the diagrams contributing to a 
$2$-point function are summed we recover the usual $\MSbar$ renormalization 
constants for the gluon and quark wave function renormalizations as well as 
that for the gluon mass parameter in the Landau gauge. This ensures that no 
rogue infrared divergence from the application of the algorithm of \cite{44} to
the masters has survived to upset the renormalizability of QCD.

\sect{Four dimensions.}

In this section we collect the results for the low momentum expansion of the
propagators in four dimensions. As a check on the computation we have
recovered the correct $\MSbar$ divergences in the Landau gauge in the presence
of a gluon mass term. This is a non-trivial check on the computation since we
are evaluating massive Feynman diagrams in a zero momentum limit. Therefore,
we have to be sure that using the approach of \cite{44} we do not incorrectly
include an infrared divergence in dimensional regularization as it is
indistinguishable from an ultraviolet one. For instance, a mismatch in the
application of the diagram subtraction process of \cite{44} could produce such 
a pole in $\epsilon$. Therefore, we note that in the four dimensional case we 
have correctly obtained the known two loop $\MSbar$ wave function
renormalization constants for the gluon and the ghost, \cite{62,63}, as well as
the Slavnov-Taylor identity for the gluon mass term. As was originally noted in
\cite{64} and subsequently rediscovered in a three loop computation in
\cite{65}, the anomalous dimension of $m$ satisfies
\begin{equation}
\gamma_m(a) ~=~ \frac{1}{2} \left[ \gamma_A(a) ~+~ \gamma_c(a) \right]
\label{sti}
\end{equation} 
in the Landau gauge where $\gamma_i(a)$ is the anomalous dimension of the 
quantity $i$, $a$~$=$~$g^2/(16\pi^2)$ and $g$ is the coupling constant. We note
that (\ref{sti}) differs by a factor of $2$ from the relation presented in
\cite{65}. This is because in \cite{65} the anomalous dimension of the gluon 
mass operator itself was determined and the renormalization constant of the 
mass operator and the mass parameter, $m$, are not equivalent but related by a
power. For clarity we note that we renormalize the gluon mass with
\begin{equation}
m_{\mbox{\footnotesize{o}}} ~=~ Z_m \, m 
\end{equation}
where ${}_{\mbox{\footnotesize{o}}}$ indicates a bare quantity, and 
$\gamma_m(a)$ is the anomalous dimension associated with the gluon mass
renormalization constant $Z_m$.

Applying the algorithm we have outlined earlier to the gluon and ghost
$2$-point functions we find that at low momentum the former is 
\begin{eqnarray}
\langle A^a_\mu(p) A^b_\nu(-p) \rangle &=& \left[ \,-~ p^2 - m^2 
\right. \nonumber \\
&& \left. ~
+ \left[ \left[ 
-~ \frac{5}{8}
+ \frac{3}{4} \ln \left( \frac{m^2}{\mu^2} \right)
\right] m^2 C_A 
\right. \right. \nonumber \\
&& \left. \left. ~~~~~~
+ \left[
-~ \frac{1}{8}
- \frac{25}{12} \ln \left( \frac{m^2}{\mu^2} \right)
- \frac{1}{12} \ln \left( \frac{p^2}{\mu^2} \right)
\right] p^2 C_A 
\right. \right. \nonumber \\
&& \left. \left. ~~~~~~
+ \left[
-~ \frac{20}{9}
+ \frac{4}{3} \ln \left( \frac{p^2}{\mu^2} \right)
\right] p^2 T_F \Nf
\right] a 
\right. \nonumber \\
&& \left. ~
+ \left[
\left[
\frac{743}{384}
-~ \frac{891}{128} s_2
+ \frac{11}{64} \zeta(2)
+ \frac{245}{96} \ln \left( \frac{m^2}{\mu^2} \right)
- \frac{53}{32} \ln^2 \left( \frac{m^2}{\mu^2} \right)
\right] m^2 C_A^2 
\right. \right. \nonumber \\
&& \left. \left. ~~~~~~
+ \left[ 
\frac{5}{24}
+ \zeta(2)
- \frac{2}{3} \ln \left( \frac{m^2}{\mu^2} \right)
+ \frac{1}{2} \ln^2 \left( \frac{m^2}{\mu^2} \right)
\right] m^2 T_F C_A \Nf 
\right. \right. \nonumber \\
&& \left. \left. ~~~~~~
+ \left[ 
-~ \frac{737}{216}
+ \frac{45}{8} s_2
- \frac{2}{9} \zeta(2)
- \frac{145}{144} \ln \left( \frac{m^2}{\mu^2} \right)
+ \frac{3}{2} \ln^2 \left( \frac{m^2}{\mu^2} \right)
\right. \right. \right. \nonumber \\
&& \left. \left. \left. ~~~~~~~~~~~
+ \frac{1}{8} \ln \left( \frac{m^2}{\mu^2} \right) 
\ln \left( \frac{p^2}{\mu^2} \right)
- \frac{5}{48} \ln \left( \frac{p^2}{\mu^2} \right)
\right] p^2 C_A^2
\right. \right. \nonumber \\
&& \left. \left. ~~~~~~
+ \left[ 
-~ \frac{139}{108}
+ 2 \zeta(2)
+ \frac{35}{9} \ln \left( \frac{m^2}{\mu^2} \right)
+ \ln^2 \left( \frac{m^2}{\mu^2} \right)
\right. \right. \right. \nonumber \\
&& \left. \left. \left. ~~~~~~~~~~~
-~ 2 \ln \left( \frac{m^2}{\mu^2} \right) \ln \left( \frac{p^2}{\mu^2} \right)
+ \frac{5}{3} \ln \left( \frac{p^2}{\mu^2} \right)
\right] p^2 T_F C_A \Nf 
\right. \right. \nonumber \\
&& \left. \left. ~~~~~~
+ \left[ 
-~ 3
+ 4 \ln \left( \frac{m^2}{\mu^2} \right)
\right] p^2 T_F C_F \Nf 
\right] a^2 ~+~ O \left( a^3; (p^2)^3 \right)
\right] P_{\mu\nu}(p) \nonumber \\
&& + 
\left[
-~ m^2
\right. \nonumber \\
&& \left. ~~~
+ \left[ \left[ 
-~ \frac{5}{8}
+ \frac{3}{4} \ln \left( \frac{m^2}{\mu^2} \right)
\right] m^2 C_A 
+ \left[
\frac{11}{24}
- \frac{1}{4} \ln \left( \frac{p^2}{m^2} \right)
\right] p^2 C_A
\right] a
\right. \nonumber \\
&& \left. ~~~
+ \left[
\left[
\frac{743}{384}
- \frac{891}{128} s_2
+ \frac{11}{64} \zeta(2)
+ \frac{245}{96} \ln \left( \frac{m^2}{\mu^2} \right)
- \frac{53}{32} \ln^2 \left( \frac{m^2}{\mu^2} \right)
\right] m^2 C_A^2 
\right. \right. \nonumber \\
&& \left. \left. ~~~~~~~
+ \left[ 
\frac{5}{24}
+ \zeta(2)
- \frac{2}{3} \ln \left( \frac{m^2}{\mu^2} \right)
+ \frac{1}{2} \ln^2 \left( \frac{m^2}{\mu^2} \right)
\right] m^2 T_F C_A N_f
\right. \right. \nonumber \\
&& \left. \left. ~~~~~~~
+ \left[ 
\frac{4967}{1152}
- \frac{459}{32} s_2
- \frac{5}{24} \zeta(2)
- \frac{53}{48} \ln \left( \frac{m^2}{\mu^2} \right)
+ \frac{3}{8} \ln \left( \frac{m^2}{\mu^2} \right) 
\ln \left( \frac{p^2}{m^2} \right)
\right. \right. \right. \nonumber \\
&& \left. \left. \left. ~~~~~~~~~~~~
- \frac{5}{16} \ln \left( \frac{p^2}{\mu^2} \right)
\right] p^2 C_A^2 
+ \left[ 
-~ \frac{5}{9}
+ \frac{1}{3} \ln \left( \frac{m^2}{\mu^2} \right)
\right] p^2 T_F C_A N_f
\right] a^2 
\right. \nonumber \\
&& \left. ~
+~ O \left( a^3; (p^2)^3 \right)
\right] L_{\mu\nu}(p)
\end{eqnarray}
where
\begin{equation}
P_{\mu\nu}(p) ~=~ \eta_{\mu\nu} ~-~ \frac{p_{\mu}p_{\nu}}{p^2} ~~~,~~~
L_{\mu\nu}(p) ~=~ \frac{p_{\mu}p_{\nu}}{p^2} 
\end{equation}
$\zeta(z)$ is the Riemann zeta function, 
$s_2$~$=$~$(2\sqrt{3}/9) \mbox{Cl}_2(2\pi/3)$ which involves the Clausen 
function $\mbox{Cl}_2(\theta)$, and $\mu$ is the renormalization scale 
introduced as a consequence of the dimensional regularization in order to 
handle the dimensionality of the coupling constant in $d$-dimensions. 
Numerically $s_2$~$=$~$0.2604341$. The order symbols are intended to indicate 
the order in which each of the coupling constant and momentum powers the 
expansion is performed to. The colour group Casimirs $C_A$, $C_F$ and $T_F$ are
given by
\begin{equation}
T^a T^a ~=~ C_F I ~~,~~ f^{acd} f^{bcd} ~=~ C_A \delta^{ab} ~~,~~
\mbox{Tr} \left( T^a T^b \right) ~=~ T_F \delta^{ab}
\end{equation}
where $T^a$ are the generators of the colour group whose structure constants
are $f^{abc}$, $I$ is the unit matrix which with $T^a$ complete the basis for
colour space. To appreciate the relative properties of the loop corrections we 
have plotted the coefficient of $(-1) m^2 a^L$ at zero momentum for 
$\Nf$~$=$~$3$ where $L$~$=$~$0$, $1$ and $2$ in Figure $1$ where
$x$~$=$~$m^2/\mu^2$. For low values of $x$ the corrections have the same
signs. 

{\begin{figure}[ht]
\hspace{4cm}
\includegraphics[width=7.6cm,height=8cm]{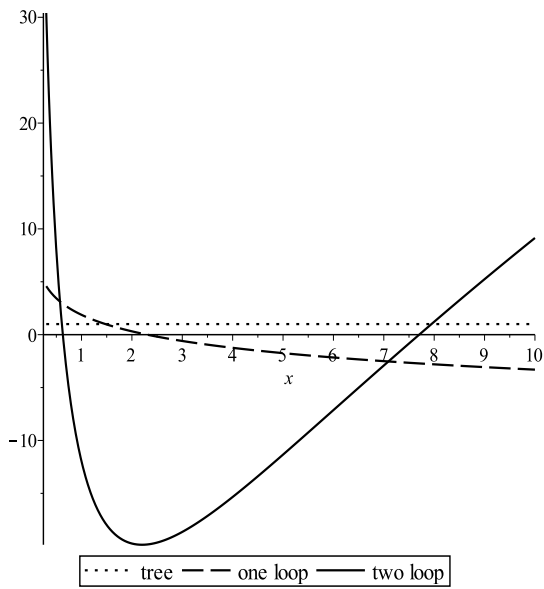}
\caption{Plots of the leading term, one loop and two loop coefficients of 
$(-1) m^2$ in the four dimensional zero momentum gluon propagator for 
$\Nf$~$=$~$3$.}
\end{figure}}

The situation for the ghost is somewhat similar though simpler since  
\begin{eqnarray}
\langle c^a(p) \bar{c}^b(-p) \rangle &=& \left[ \,-~ 1  
+ \left[
\frac{5}{8}
- \frac{3}{4} \ln \left( \frac{m^2}{\mu^2} \right)
\right] C_A a
\right. \nonumber \\
&& \left. ~
+ \left[
\left[
-~ \frac{893}{384}
+ \frac{891}{128} s_2
- \frac{11}{64} \zeta(2)
- \frac{155}{96} \ln \left( \frac{m^2}{\mu^2} \right)
+ \frac{35}{32} \ln^2 \left( \frac{m^2}{\mu^2} \right)
\right] C_A^2
\right. \right. \nonumber \\
&& \left. \left. ~~~~~
+ \left[
-~ \frac{5}{24}
- \zeta(2)
+ \frac{2}{3} \ln \left( \frac{m^2}{\mu^2} \right)
- \frac{1}{2} \ln^2 \left( \frac{m^2}{\mu^2} \right)
\right] T_F C_A \Nf
\right] a^2 
\right. \nonumber \\
&& \left. ~~~~~
\,+~ O(a^3) \right] p^2 ~+~ O \left( (p^2)^2 \right) ~.
\end{eqnarray}
One interesting feature of both $2$-point functions resides in the one loop
leading order terms which are very similar. For instance, if we were in a
Gribov scenario then there is a gap equation for the corresponding mass 
parameter deriving from the horizon condition defining the boundary of the 
first Gribov region, \cite{14}. This condition when imposed means that the 
ghost propagator enhances and, moreover, satisfies the Kugo-Ojima confinement 
criterion, \cite{45,46}. However, if one wished to impose a Kugo-Ojima type 
confinement criterion on the Faddeev-Popov ghost propagator in order to have 
ghost enhancement it might appear that at the same stroke the gluon propagator 
would be massless at zero momentum. A closer examination of the actual signs 
clearly indicates, however, that such a scenario does not occur. Setting such a
pseudo-Kugo-Ojima criterion, for example, merely retains a non-zero value for 
the gluon propagator at zero momentum or in other words a frozen propagator 
consistent with the decoupling scenario, \cite{10,11}, favoured by lattice 
data, \cite{1,2,3,4,5,6,7,8,9}. The Faddeev-Popov ghost would then enhance but 
this is not observed in lattice data when there is a non-zero frozen gluon
propagator.

\sect{Three dimensions.}

We have repeated the analysis for the three dimensional case. The procedure is
the same with the main difference being the use of different master integrals 
after the application of the Laporta reduction. Therefore, we have adapted our 
{\sc Form} code to accommodate this variation. However, there are some novel 
features in three dimensions compared with the previous section. For instance, 
the theory is superrenormalizable and hence not every Green's function is 
divergent. It transpires that to two loops both the gluon and ghost $2$-point 
functions are finite when there is a non-zero gluon mass term in the
Lagrangian. Therefore unlike the four dimensional case when the divergences for
these Green's functions are known and thereby provide an internal check on the 
computations, there is no such check in this case. However, the use of 
predominantly the same code aside from master integrals should ensure the 
correctness of our results in this instance. Another aspect of the 
superrenormalizability relates to the dimensionality of the coupling constant 
which is no longer dimensionless. Instead in three dimensions $a$ has mass 
dimension one. One point which may arise concerning the applicability of the 
approach to the three dimensional case relates to the work of \cite{66}. In
\cite{66} massless gauge theories were examined at two loops due to the 
potential breakdown of perturbation theory. However, it was shown that this was
not the case. Here the corresponding situation does not arise since the 
presence of the mass obviates this issue.

For the gluon $2$-point function we find that to two loops the low momentum
structure is 
\begin{eqnarray}
\langle A^a_\mu(p) A^b_\nu(-p) \rangle &=& \left[ \,-~ p^2 - m^2 
+ \left[ \frac{1}{64} \sqrt{p^2} C_A - \frac{1}{8} \sqrt{p^2} T_F \Nf
- \frac{1}{6\pi} m C_A
+ \frac{9}{40\pi} \frac{p^2}{m} C_A \right] a
\right. \nonumber \\
&& \left. ~
+ \left[ \frac{1}{192\pi} \frac{\sqrt{p^2}}{m} C_A^2
- \frac{1}{24\pi} \frac{\sqrt{p^2}}{m} C_A T_F \Nf
\right. \right. \nonumber \\
&& \left. \left. ~~~~~
+ \left[
-~ \frac{121}{1536}
- \frac{1}{12} \ln(2)
+ \frac{21}{256} \ln(3)
\right] \frac{C_A^2}{\pi^2}
+ \frac{1}{48\pi^2} C_A T_F \Nf
\right. \right. \nonumber \\
&& \left. \left. ~~~~~
+ \left[
-~ \frac{11951}{691200}
+ \frac{9}{1280} \ln(3)
+ \frac{1}{1280} \ln \left( \frac{p^2}{m^2} \right)
\right] \frac{p^2}{m^2} \frac{C_A^2}{\pi^2}
- \frac{1}{16384} \frac{p^2}{m^2} C_A^2
\right. \right. \nonumber \\
&& \left. \left. ~~~~~
+ \left[
\frac{139}{1800\pi^2}
- \frac{1}{80\pi^2} \ln \left( \frac{p^2}{m^2} \right)
\right] \frac{p^2}{m^2} C_A T_F \Nf
- \frac{1}{512} \frac{p^2}{m^2} C_A T_F \Nf
\right. \right. \nonumber \\
&& \left. \left. ~~~~~
- \frac{1}{16\pi^2} \frac{p^2}{m^2} C_F T_F \Nf
+ \frac{1}{256} \frac{p^2}{m^2} C_F T_F \Nf
\right] a^2 ~+~ O \left( a^3 ; (p^2)^{\threehalves} \right)
\right] P_{\mu\nu}(p) \nonumber \\
&& + 
\left[ \,-~ m^2
+ \left[ \frac{1}{32} \sqrt{p^2} C_A 
- \frac{1}{6\pi} m C_A
- \frac{1}{30\pi} \frac{p^2}{m} C_A \right] a
\right. \nonumber \\
&& \left. ~
+ \left[ \frac{1}{96\pi} \frac{\sqrt{p^2}}{m} C_A^2
+ \left[
-~ \frac{121}{1536}
- \frac{1}{12} \ln(2)
+ \frac{21}{256} \ln(3)
\right] \frac{C_A^2}{\pi^2}
+ \frac{1}{48\pi^2} C_A T_F \Nf
\right. \right. \nonumber \\
&& \left. \left. ~~~~~
+ \left[
-~ \frac{1783}{115200}
- \frac{5}{96} \ln(2)
+ \frac{13}{320} \ln(3)
+ \frac{1}{960} \ln \left( \frac{p^2}{m^2} \right)
\right] \frac{p^2}{m^2} \frac{C_A^2}{\pi^2}
\right. \right. \nonumber \\
&& \left. \left. ~~~~~
- \frac{3}{4096} \frac{p^2}{m^2} C_A^2
+ \left[
-~ \frac{31}{3600\pi^2}
+ \frac{1}{240\pi^2} \ln \left( \frac{p^2}{m^2} \right)
\right] \frac{p^2}{m^2} C_A T_F \Nf
\right] a^2 
\right. \nonumber \\
&& \left. 
+~ O \left( a^3 ; (p^2)^{\threehalves} \right) \right] L_{\mu\nu}(p) ~.  
\end{eqnarray}
In terms of structure for these corrections compared to the four dimensional
case the role $s_2$ played is taken now by the presence of $\ln(2)$ and 
$\ln(3)$. For the ghost $2$-point function the situation is somewhat simpler 
since 
\begin{eqnarray}
\langle c^a(p) \bar{c}^b(-p) \rangle &=& \left[ \,-~ 1  
+ \frac{C_A}{6\pi m} a \right. \nonumber \\
&& \left. ~
+ \left[ \left[
\frac{235}{4608}
+ \frac{1}{12} \ln(2)
- \frac{21}{256} \ln(3)
\right] \frac{C_A^2}{\pi^2 m^2} 
- \frac{1}{48} \frac{C_A T_F \Nf}{\pi^2 m^2} 
\right] a^2 ~+~ O(a^3) \right] p^2 \nonumber \\
&& +~ O \left( (p^2)^2 \right) ~.
\end{eqnarray}
Moreover, we observe that there is a similar structure to the four dimensional
case. Within the one loop ghost $2$-point function one can observe the origins
of the Gribov mass gap equation and the same comments apply in relation to what
was stated for the Kugo-Ojima criterion earlier. Equally the situation is 
blurred at two loops since the mixed mass scale master integrals can not be 
adduced within the above results.  

\sect{Discussion.}

We conclude with brief remarks. In analysing the two loop corrections to the
gluon and ghost propagators in this effective theory of Gribov copies,
\cite{30,31,32}, we have provided useful values for the approach to zero 
momentum. This will be important for the programme proposed in \cite{31} which
indicated that the exact evaluation of the two loop corrections would be
possible. For instance, having {\em independent} computations, albeit in the
low momentum limit, is crucial to establishing the correctness of an exact
result. However, the determination of the two loop correction to where the
gluon propagator freezes emerges from this calculation. Indeed there are
intriguing similarities to the structure of the corrections when compared to
propagators evaluated in the Gribov-Zwanziger construction. For instance,
a pseudo-mass gap appears to be present which is similar in terms of numerical
values to the gap equation for the Gribov mass. 

\vspace{1cm}
\noindent
{\bf Acknowledgements.} The author thanks Dr M. Tissier for useful discussions.

\appendix

\sect{Master integrals.}

In this appendix we give the $\epsilon$ expansion for several master integrals 
which were required in three and four dimensions. We have applied the method of 
\cite{44} and have reproduced the expressions given for the various integrals 
in Table $1$ of \cite{44}. As the {\sc Reduze} algorithm produces additional 
master integrals for our particular computation we note that several of these 
are
\begin{eqnarray} 
I_{m0m00}(1,1,1,0,1) &=& \frac{1}{2\epsilon^2}
+ \left[ \frac{1}{2} - \ln \left( \frac{m^2}{\mu^2} \right) \right] 
\frac{1}{\epsilon} 
+ \frac{1}{2} - \ln \left( \frac{m^2}{\mu^2} \right) 
+ \ln^2 \left( \frac{m^2}{\mu^2} \right) 
+ \frac{3}{2} \zeta(2) \nonumber \\
&& + \left[ - \frac{1}{6\epsilon} - \frac{1}{3}  
+ \frac{1}{3} \ln \left( \frac{m^2}{\mu^2} \right) \right] \frac{p^2}{m^2}
\nonumber \\
&& + \left[ \frac{1}{60\epsilon} + \frac{11}{120} 
- \frac{1}{30} \ln \left( \frac{m^2}{\mu^2} \right) \right] \frac{(p^2)^2}{m^4}
\nonumber \\
&& + \left[ - \frac{1}{420\epsilon} - \frac{1}{45}  
+ \frac{1}{210} \ln \left( \frac{m^2}{\mu^2} \right) \right] 
\frac{(p^2)^3}{m^6} ~+~ O \left( \frac{(p^2)^4}{m^8} \right) \nonumber \\
I_{mm00m}(1,1,1,0,1) &=& \frac{1}{2\epsilon^2}
+ \left[ \frac{3}{2} - \ln \left( \frac{m^2}{\mu^2} \right) \right] 
\frac{1}{\epsilon} 
+ \frac{7}{2} - \frac{27}{2} s_2 
- 3 \ln \left( \frac{m^2}{\mu^2} \right) 
+ \ln^2 \left( \frac{m^2}{\mu^2} \right) \nonumber \\
&& +~ \frac{1}{2} \zeta(2) 
+ \left[ - \frac{1}{2\epsilon} - \frac{3}{4} 
+ \ln \left( \frac{m^2}{\mu^2} \right) \right] \frac{p^2}{m^2}
\nonumber \\
&& + \left[ \frac{1}{6\epsilon} + \frac{1}{2} 
- \frac{1}{3} \ln \left( \frac{m^2}{\mu^2} \right) \right] \frac{(p^2)^2}{m^4}
\nonumber \\
&& + \left[ - \frac{1}{12\epsilon} - \frac{113}{360} 
+ \frac{1}{6} \ln \left( \frac{m^2}{\mu^2} \right) \right] 
\frac{(p^2)^3}{m^6} ~+~ O \left( \frac{(p^2)^4}{m^8} \right) \nonumber \\
I_{mmm0m}(1,1,1,0,1) &=& \frac{1}{2\epsilon^2}
+ \left[ \frac{1}{2} - \ln \left( \frac{m^2}{\mu^2} \right) \right] 
\frac{1}{\epsilon} 
+ \frac{1}{2} - \frac{9}{2} s_2 
- \ln \left( \frac{m^2}{\mu^2} \right) 
+ \ln^2 \left( \frac{m^2}{\mu^2} \right) 
+ \frac{1}{2} \zeta(2) \nonumber \\
&& + \left[ - \frac{1}{6\epsilon} - \frac{2}{9} + s_2 
+ \frac{1}{3} \ln \left( \frac{m^2}{\mu^2} \right) \right] \frac{p^2}{m^2}
\nonumber \\
&& + \left[ \frac{1}{60\epsilon} + \frac{247}{3240} - \frac{2}{9} s_2 
- \frac{1}{30} \ln \left( \frac{m^2}{\mu^2} \right) \right] \frac{(p^2)^2}{m^4}
\nonumber \\
&& + \left[ - \frac{1}{420\epsilon} - \frac{113}{4860} + \frac{2}{27} s_2 
+ \frac{1}{210} \ln \left( \frac{m^2}{\mu^2} \right) \right] 
\frac{(p^2)^3}{m^6} ~+~ O \left( \frac{(p^2)^4}{m^8} \right) \nonumber \\
I_{mm00m}(1,1,1,1,1) &=& \left[ \frac{27}{2} s_2 - \zeta(2) 
- \left[ \frac{3}{2} + \frac{27}{4} s_2 - \frac{3}{2} \zeta(2) \right] 
\frac{p^2}{m^2}
+ \left[ \frac{47}{18} - \frac{4}{3} \zeta(2) \right] \frac{(p^2)^2}{m^4}
\right. \nonumber \\
&& \left. ~
- \left[ \frac{1703}{720} - \frac{27}{8} s_2 - \frac{3}{4} \zeta(2) \right] 
\frac{(p^2)^3}{m^6}
\right] \frac{1}{m^2} ~+~ O \left( \frac{(p^2)^4}{m^{10}} \right) \nonumber \\
I_{mmmmm}(1,1,1,1,1) &=& \left[ 3 s_2 
- \left[ \frac{2}{27} + \frac{1}{6} s_2 \right] \frac{p^2}{m^2}
+ \left[ \frac{29}{972} - \frac{1}{27} s_2 \right] \frac{(p^2)^2}{m^4}
\right. \nonumber \\
&& \left. ~
- \left[ \frac{367}{29160} - \frac{11}{324} s_2 \right] \frac{(p^2)^3}{m^6}
\right] \frac{1}{m^2} ~+~ O \left( \frac{(p^2)^4}{m^{10}} \right)
\end{eqnarray}
in four dimensions.

In three dimensions similar masters are required but in contrast to four
dimensions they do not have poles in $\epsilon$. For comparison we record the
same masters are
\begin{eqnarray} 
I_{m0m00}(1,1,1,0,1) &=& \left[ \frac{1}{32}
- \frac{5}{576} \frac{p^2}{m^2}
+ \frac{1}{300} \frac{(p^2)^2}{m^4}
- \frac{151}{94080} \frac{(p^2)^3}{m^6}
\right] \frac{1}{\pi^2 m^2} \nonumber \\
&& +~ O \left( \frac{(p^2)^4}{m^{10}} \right) \nonumber \\ 
I_{mm00m}(1,1,1,0,1) &=& \left[ -~ \frac{1}{16} \ln \left( \frac{2}{3} \right)
+ \left[ - \frac{1}{384} + \frac{1}{48} \ln \left( \frac{2}{3} \right) \right]
\frac{p^2}{m^2}
\right. \nonumber \\
&& \left. ~ 
+ \left[ \frac{9}{5120} - \frac{1}{80} \ln \left( \frac{2}{3} \right) \right]
\frac{(p^2)^2}{m^4}
\right. \nonumber \\
&& \left. ~ 
+ \left[ -~ \frac{55}{43008} + \frac{1}{112} \ln \left( \frac{2}{3} \right) 
\right]
\frac{(p^2)^3}{m^6}
\right] \frac{1}{\pi^2 m^2} 
~+~ O \left( \frac{(p^2)^4}{m^{10}} \right) \nonumber \\ 
I_{mmm0m}(1,1,1,0,1) &=& \left[ \frac{1}{96}
- \frac{5}{3888} \frac{p^2}{m^2}
+ \frac{323}{1555200} \frac{(p^2)^2}{m^4}
- \frac{20789}{548674560} \frac{(p^2)^3}{m^6}
\right] \frac{1}{\pi^2 m^2} \nonumber \\
&& +~ O \left( \frac{(p^2)^4}{m^{10}} \right) \nonumber \\
I_{mm00m}(1,1,1,1,1) &=& \left[ \frac{1}{16} \ln \left( \frac{4}{3} \right)
+ \left[ \frac{1}{192} - \frac{1}{12} \ln \left( \frac{4}{3} \right) \right]
\frac{p^2}{m^2}
\right. \nonumber \\
&& \left. ~ 
+ \left[ \frac{13}{7680} + \frac{13}{240} \ln \left( \frac{4}{3} \right) 
\right]
\frac{(p^2)^2}{m^4}
\right. \nonumber \\
&& \left. ~ 
+ \left[ -~ \frac{1531}{107520} - \frac{1}{210} \ln \left( \frac{4}{3} \right) 
\right]
\frac{(p^2)^3}{m^6}
\right] \frac{1}{\pi^2 m^4} \nonumber \\
&& +~ O \left( \frac{(p^2)^4}{m^{12}} \right) \nonumber \\ 
I_{mmmmm}(1,1,1,1,1) &=& \left[ \frac{1}{576}
- \frac{1}{1944} \frac{p^2}{m^2}
+ \frac{1451}{11197440} \frac{(p^2)^2}{m^4}
- \frac{2941}{94058496} \frac{(p^2)^3}{m^6}
\right] \frac{1}{\pi^2 m^4} \nonumber \\
&& +~ O \left( \frac{(p^2)^4}{m^{12}} \right) ~.
\end{eqnarray} 
The expressions for the four dimensional case should be useful for other
problems.

\end{document}